# Design of a Fast Crowbar Protection Circuit for Single Event Effect Testing

C. Burt, J. D'Amico, J. Van Grinsven, T. Aldaz, and N. Dodds
Sandia National Laboratories, Albuquerque, NM, USA

***Abstract:***
We describe the design of a custom crowbar circuit that quickly removes power when a programmable current threshold is exceeded. This protects devices from permanent damage during single event latchup/burnout tests.

**Corresponding & Presenting Author:**

Collin Burt
Sandia National Laboratories, Albuquerque, NM 87123 (USA)
Phone: (505) 845-1201, Email: cjburt@sandia.gov

**Session Preference:** Data Workshop

**Presentation Preference:** Poster

## I. Introduction

Some single event effect (SEE) tests for high-current failure modes like latchup or burnout are simple screening tests to look for red flags, after which any failing parts are rejected. However, often more involved single event latchup (SEL) or single event burnout (SEB) testing is required to measure SEL/SEB cross sections, or to establish safe operating areas (as a function of bias voltage, linear energy transfer, temperature, etc.), or to locate sensitive regions via focused laser testing. When the latchup/burnout is destructive, meaning when it permanently damages the device, then it becomes very difficult and labor intensive to complete such tests, requiring very large numbers of samples and frequent sample replacement resulting in lost time and increased cost. This work presents the design of a custom crowbar circuit that quickly removes power from a device under test (DUT) once a programmable current or voltage threshold is exceeded. That design is then copied three times and implemented on a printed circuit board (PCB), Figure 1, with different program components for protecting 0.8V, 1.8V, and 3.3V rails. This circuitry protects the DUT from SEL/SEB much better and on shorter timescales than alternative methods like setting current limits on power supplies. Therefore, the crowbar circuit minimizes the chance of the high current mode causing permanent damage and enables more testing to be performed on a given sample. The decision to design a custom crowbar circuit was made after performing a review of commercial alternatives that were deemed insufficient for various reasons such as speed of operation, crowbar topology, etc.

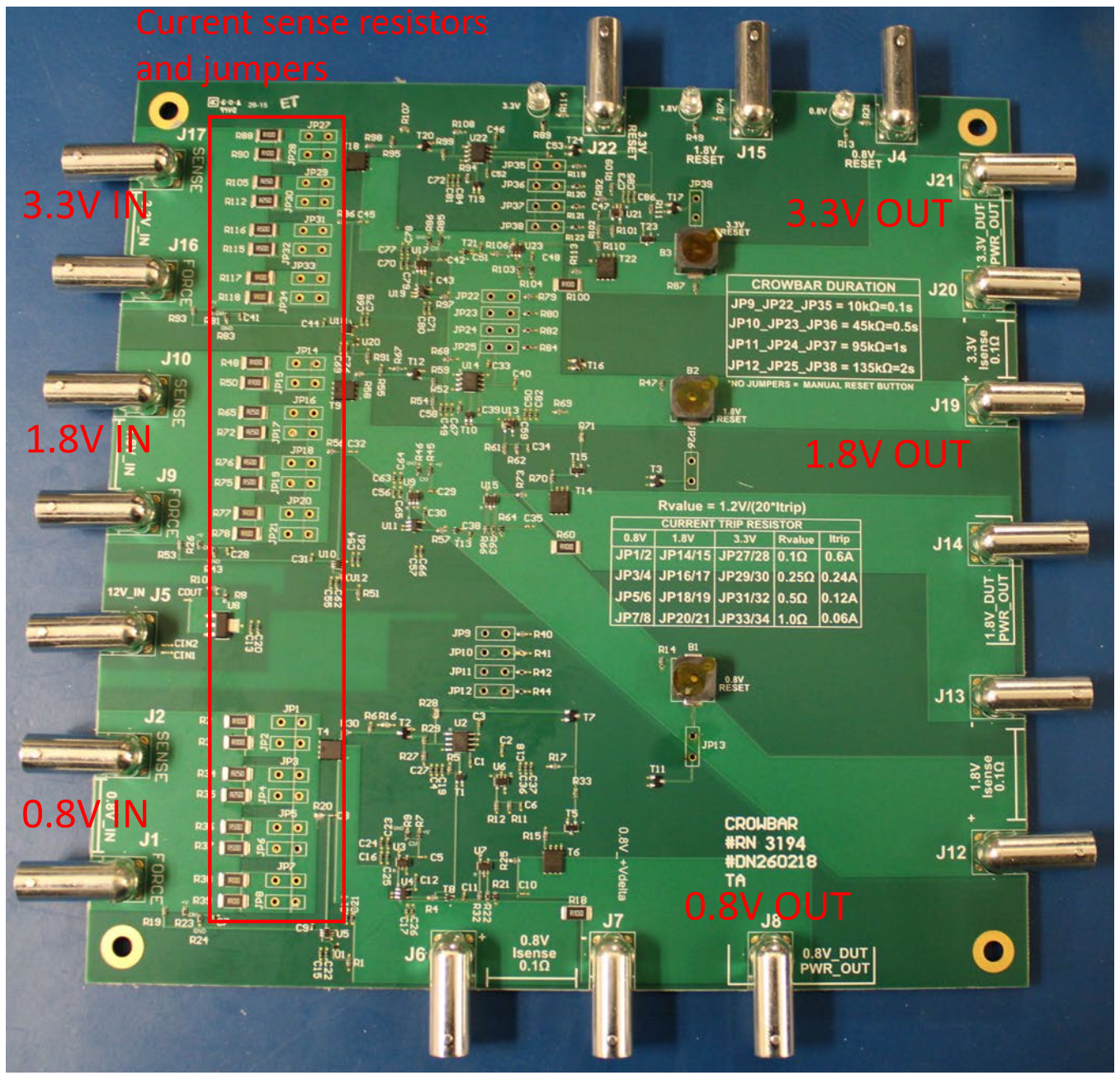

Figure 1. Crowbar PCB programmed to support 0.8V, 1.8V, and 3.3V supply rails.

## II. Crowbar Circuit Design

### A. Crowbar Circuit

The crowbar circuit is meant to be placed between the power supply and the device under test (DUT). If an overvoltage or overcurrent condition is detected, then the crowbar circuit will actuate and rapidly cut power to the DUT by breaking the connection to the power supply and connecting the DUT supply input to ground. The following sections describe the subcircuit designs that culminate in the final crowbar circuit implementation. All simulations are run in Texas Instruments' TI-TINA software [1]. The VCC node common to many of the schematics is a 5V supply rail for powering the integrated circuits (ICs) on the crowbar PCB.

### B. Crowbar Transistors

The primary transistors involved in the crowbar action are shown in Figure 2. In this case, the DUT is the parallel combination of RLOAD and CLOAD. 0.8V is supplied to the load through a resistor and a transistor. $R_{sense}$ provides a small voltage drop that is used by other circuitry to sense an overcurrent event. T-PASS is a pass transistor that is normally conducting but shuts off when the crowbar is initiated, breaking the connection between the DUT and the power supply (a common ground connection is always present between the power supply and the DUT). The other transistor, T-SHUNT, is used to discharge stored energy in the output capacitance (decoupling capacitors and any capacitance on the DUT or internal to the DUT packaging) to reduce current draw through the DUT. This can result in large currents being pulled through T-SHUNT. To limit this current, it may be desirable to move the bulk of the DUT's decoupling capacitors from the VOUT node to the drain of the T-PASS transistor. When using both force and sense on a power supply (i.e., 4-wire Kelvin connection), the supply sense can be connected from ground to the drain of T-PASS to prevent the voltage from climbing when the connection to the DUT is broken. If this technique is used, then it may be desirable to set the power supply voltage slightly higher to account for voltage drops from the T-PASS transistor.

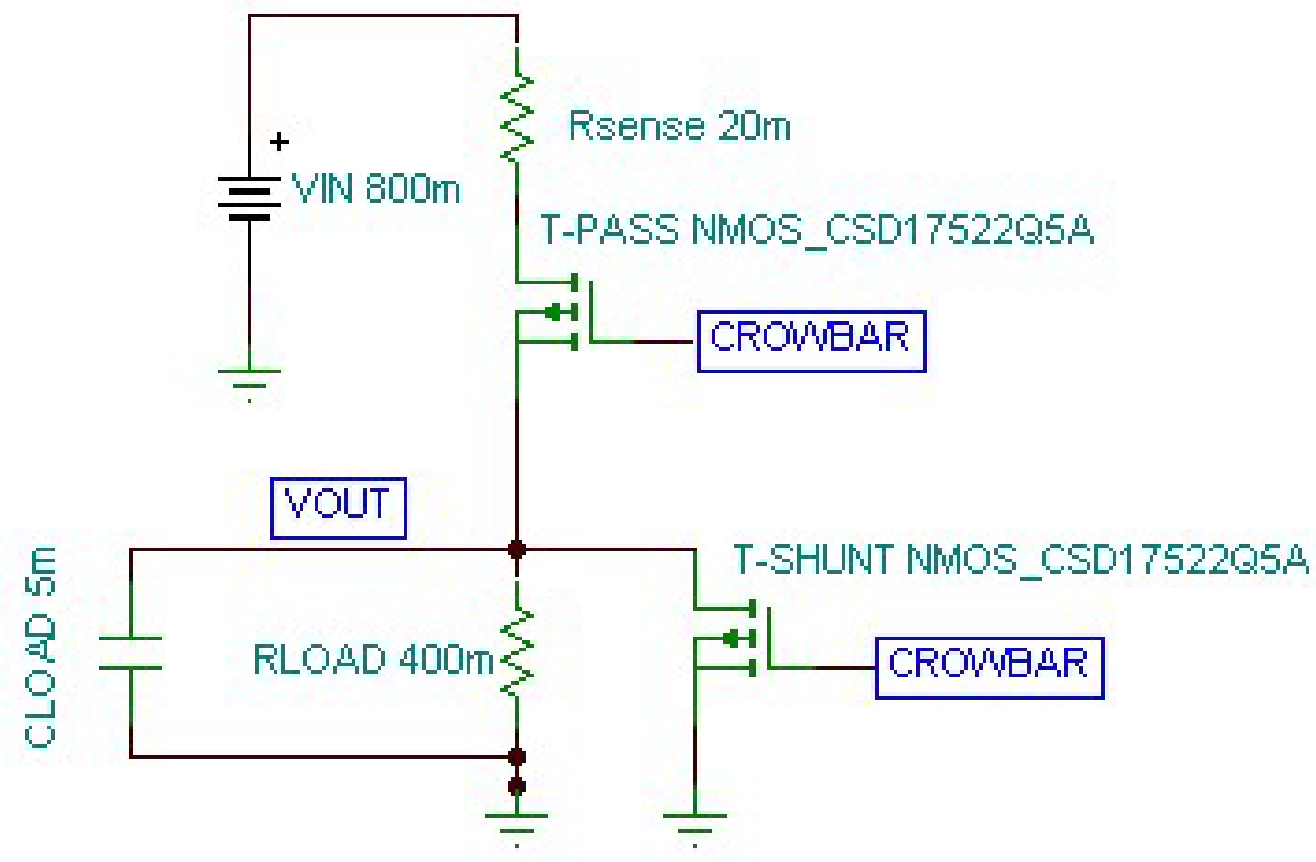

Figure 2. Crowbar output circuitry schematic.

### C. Overvoltage and Overcurrent Protection

To sense an overvoltage or overcurrent event, the circuits shown in Figure 3 and Figure 4 are used, respectively. Overvoltage detection is accomplished by monitoring the input voltage through a voltage divider and comparing it with a reference voltage. This is beneficial for protecting against issues that may arise from incorrectly setting the supply voltage, improper use of supply force and sense, or potential oscillations in the power supply output as a result of unstable operating conditions. The overvoltage threshold is set by adjusting the voltage divider ratios. Overcurrent protection is designed according to [2], a Texas Instruments application note. A voltage is produced across $R_{sense}$ when current is passed through it. That voltage signal is amplified by the IN185A1 [3], a high-precision current sense amplifier, and sent to a comparator with a built-in reference. If the current is high enough, then the reference signal is exceeded and the output changes logic states. The overcurrent threshold is set by the $R_{sense}$ resistor. The procedure for calculating the value of $R_{sense}$ is outlined in [2] and is based on the gain of the amplifier (the INA185A1 in this case), the reference voltage of the comparator, and the desired threshold current. For this circuit, several $R_{sense}$ values are calculated to account for a range of threshold currents, and they are implemented on the printed circuit board with jumpers to offer a quick and easy way to set or modify the current threshold.

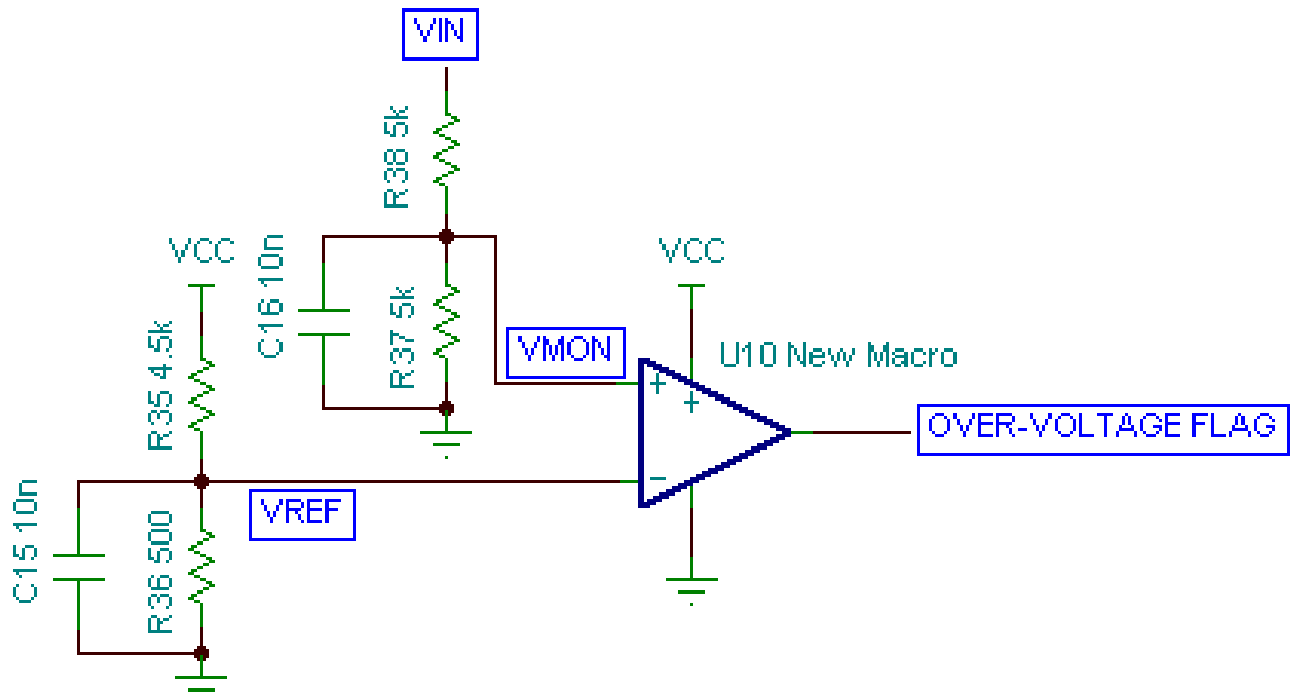

Figure 3. Overvoltage detection circuit schematic.

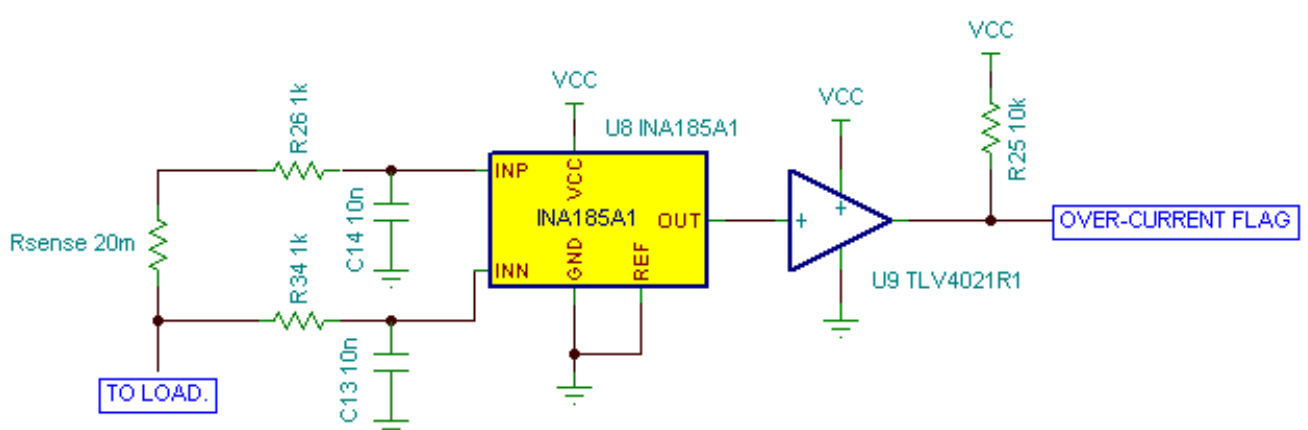

Figure 4. Overcurrent detection circuit schematic.

## D. 555 Timer

While it is possible to use the output of either the voltage or current detection circuits to drive the crowbar circuitry directly, that could result in oscillations that rapidly power cycle the part. To mitigate this concern, both signals are first fed into an OR gate and sent to a 555 Timer in a monostable circuit configuration [3], Figure 5, used to introduce a fixed delay and initiate the crowbar action. The delay is easily adjustable by changing resistor and capacitor values, or it can be overridden using switches to enable an operating mode where the circuit must always be reset manually and can hold its state for any duration. The output of the 555 Timer is sent through several transistors serving as buffers before driving the crowbar transistors. Furthermore, the output of the 555 Timer is a good place to add visual and audible indicators to alert the user that crowbar has occurred. This can be accomplished using the circuit shown in Figure 6 that compares the output of the 555 Timer with a reference voltage to turn on an LED and buzzer whenever the crowbar circuit is actuated.

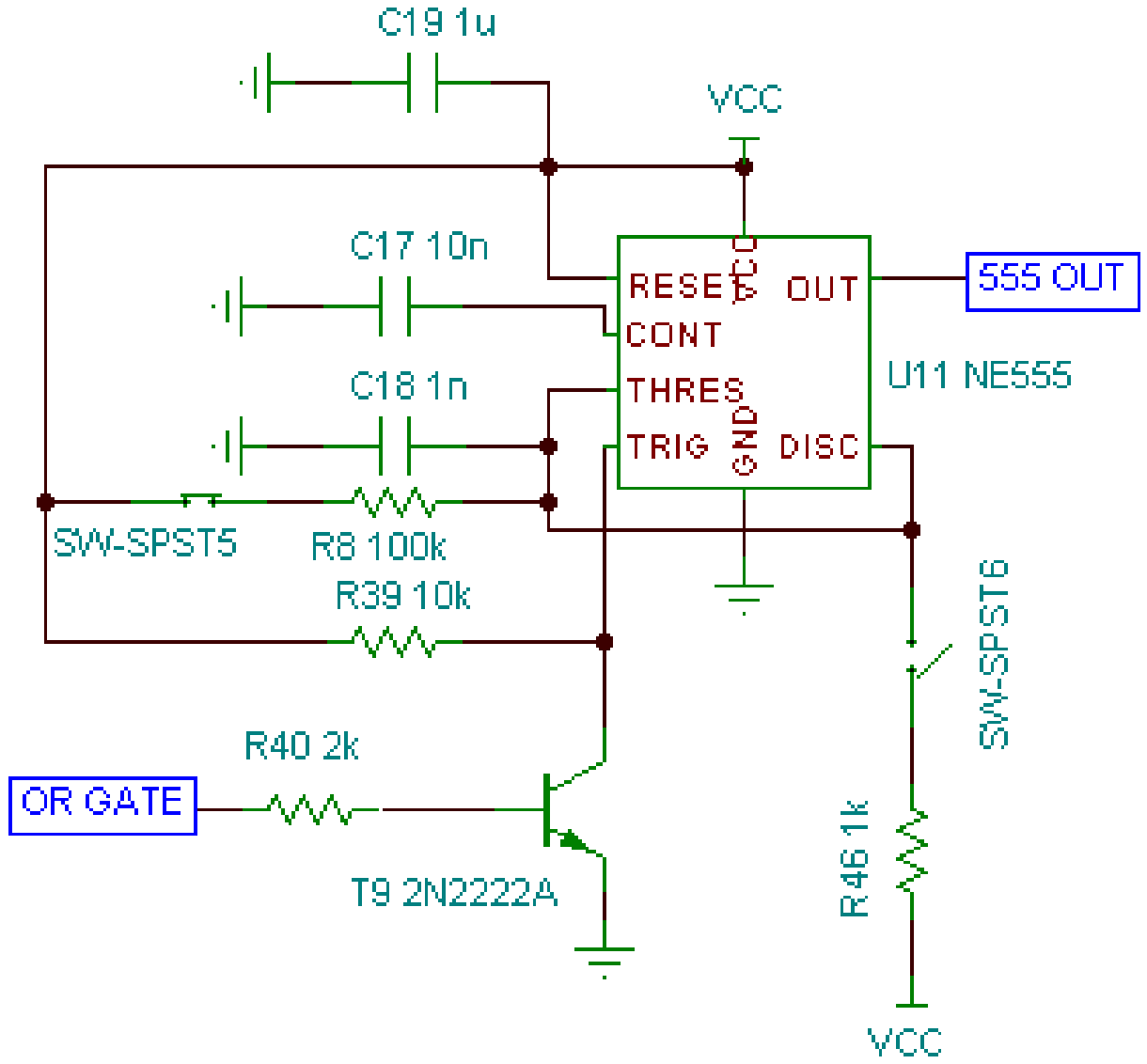

Figure 5. 555 Timer circuit schematic.

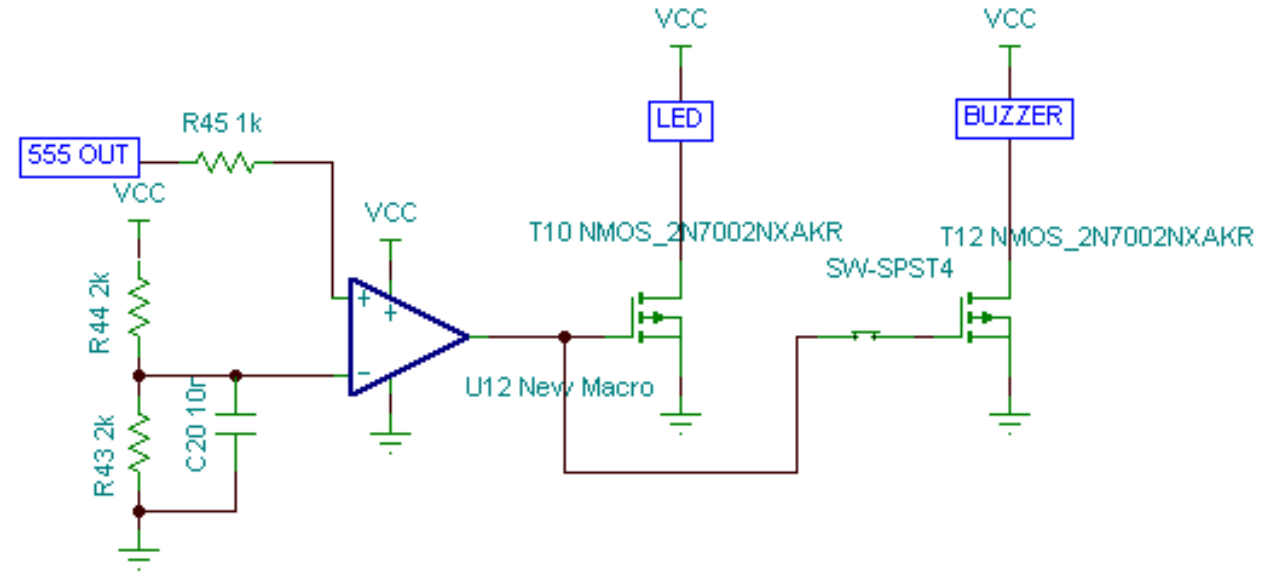

Figure 6. Crowbar indicator circuit schematic.

## E. Powerup Delay Circuit

During powerup, it is common to have a large inrush current flow from the power supply to charge up any load capacitance on the DUT and/or PCB. There are many cases where that inrush current will be large enough to prematurely trigger the overcurrent detection circuitry, and that may result in oscillations and prevent the output from charging. To address this situation, the circuit in Figure 7 is used. It compares the output voltage with a programmable reference voltage and then drives a transistor through an RC-filter that introduces a delay. On startup, the comparator output is high, and the transistor is conducting. Figure *9* shows how that transistor connects to the 555 Timer circuitry. The delay prevents any logic-high signals from the OR gate monitoring the overvoltage and overcurrent circuits from turning on the bipolar junction transistor (BJT) and triggering the 555 Timer before the output is charged. When the output is charged, then the output of the comparator turns off, enabling triggering of the 555 Timer from overvoltage or overcurrent events.

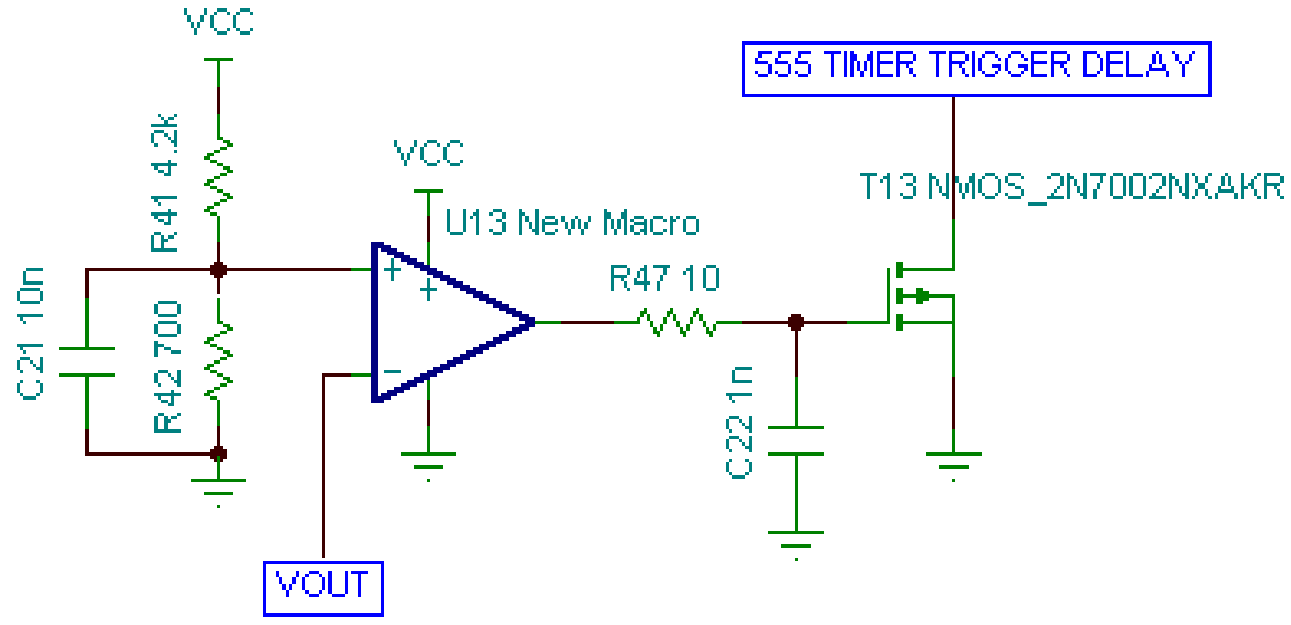

Figure 7. Powerup delay circuit schematic.

### F. *Final Circuit Design*

The subcircuits described in the previous sections are combined with additional circuitry to create the single-channel crowbar circuit shown in Figure 8. This design provides the necessary crowbar circuitry to protect a single voltage rail. It is copied three times and implemented on a PCB with different program components for protecting 0.8V, 1.8V, and 3.3V rails. A 12V supply is also used to provide higher voltages for driving the high-side FETs and a linear regulator is used to convert the 12V input into 5V for powering the various integrated circuits. Final simulation results for the 0.8V crowbar circuit are shown in Figure 10. The topmost signals show simulated overcurrent and overvoltage events, each of which cause the 555 Timer output to latch for a specified delay that initiates the crowbar transistors and cuts the output voltage/current.

The PCBs arrived and were tested on a benchtop to verify functionality. To emulate a high-current event, power is supplied at the input of the crowbar circuit, and an uncharged capacitor is connected to the DUT output. This results in a burst of excess current that trips the overcurrent protection circuitry and initiates the crowbar function. Test results showing the delay from the output disruption on a 3.3V rail to power being cut by the crowbar circuit are shown in Figure 11. This verifies that the protection circuitry is working as designed. It is important to note that this test was conducted with no load and that a large amount of decoupling capacitance at the output will increase the power down time as it takes additional time to discharge the capacitors. SEL/SEB experiments are being conducted at the time of writing to verify that the circuitry can protect an advanced-node complimentary metal-oxide-semiconductor (CMOS) circuit in an actual test environment.

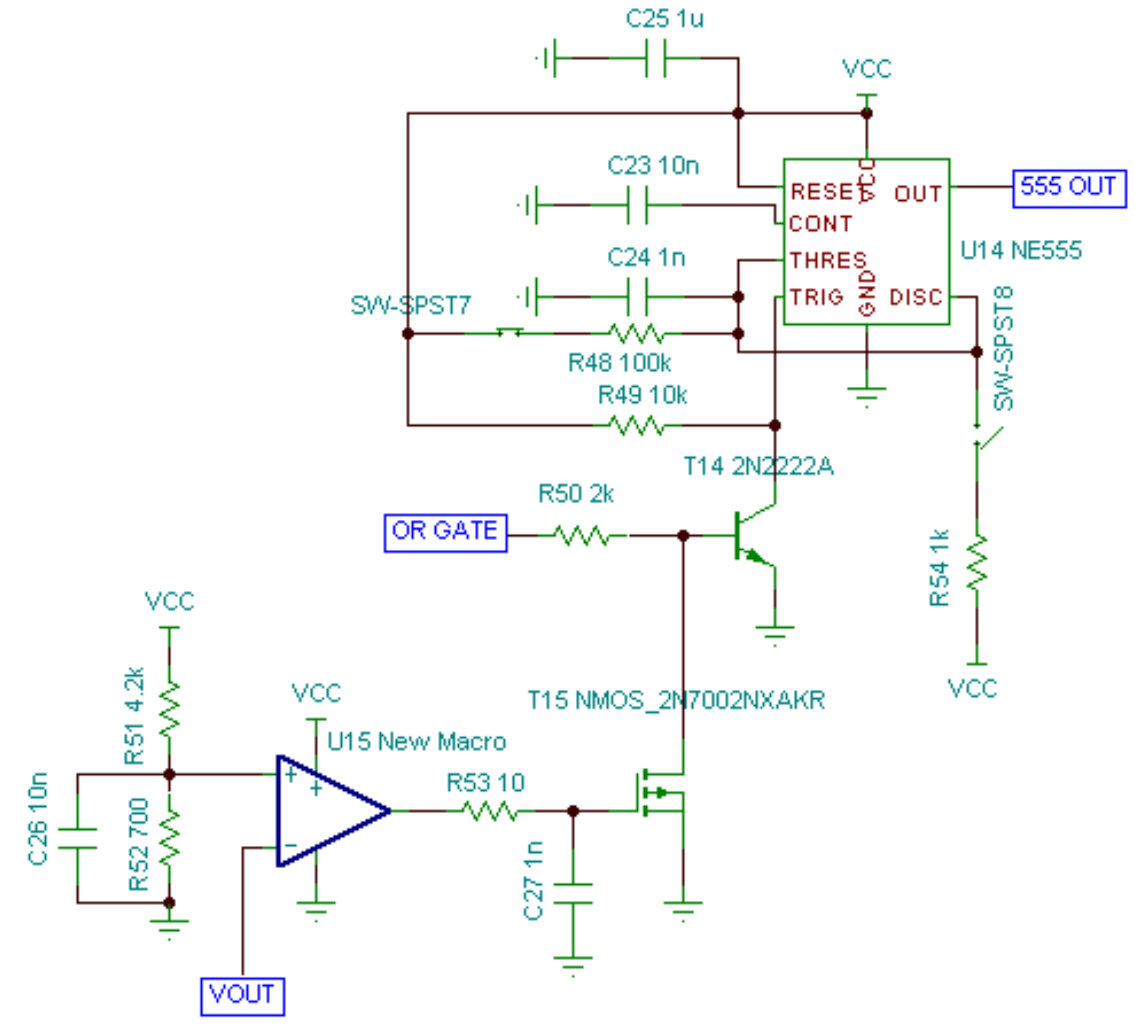

Figure 9. Powerup delay circuitry connected to the 555 Timer trigger.

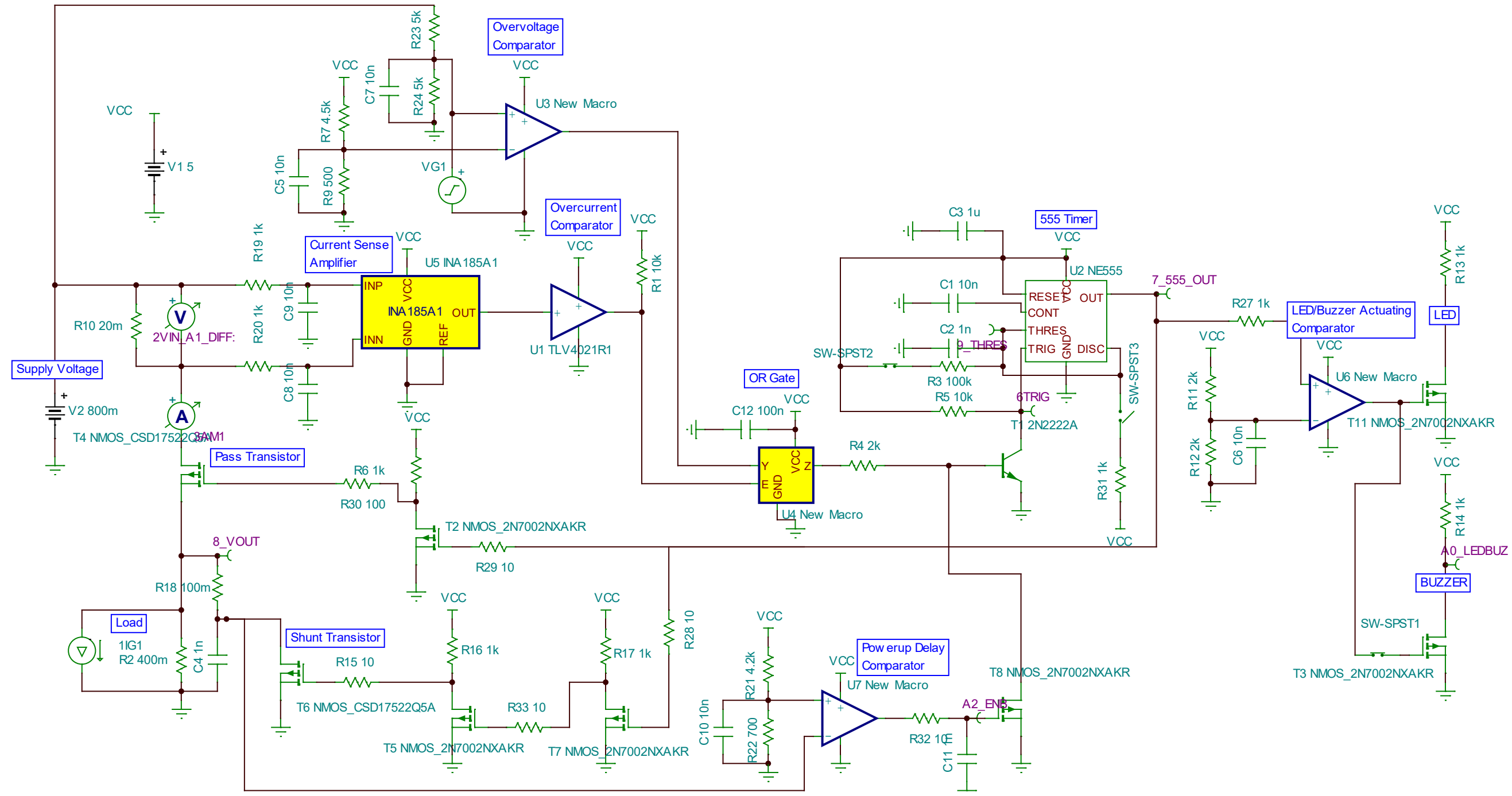

Figure 8. Single-channel crowbar schematic.

## III. Conclusions

This work describes the design and function of a fast crowbar circuit. It is designed to protect DUTs against high-current latchup/burnout during single event effects testing. It has been tested on the benchtop to demonstrate general functionality and single event testing is being performed at the time of writing. This circuit is expected to result in significant time and cost savings as a single device under test can be protected against burnout and subjected to non-destructive SEL repeatedly rather than having to continuously swap the DUT as parts are damaged or destroyed. The conference presentation will also include experimental data demonstrating that SEL was destructive in a certain complementary metal-oxide-semiconductor (CMOS) part without the crowbar circuit, and that using the crowbar circuit prevented destruction in subsequent SEL tests.

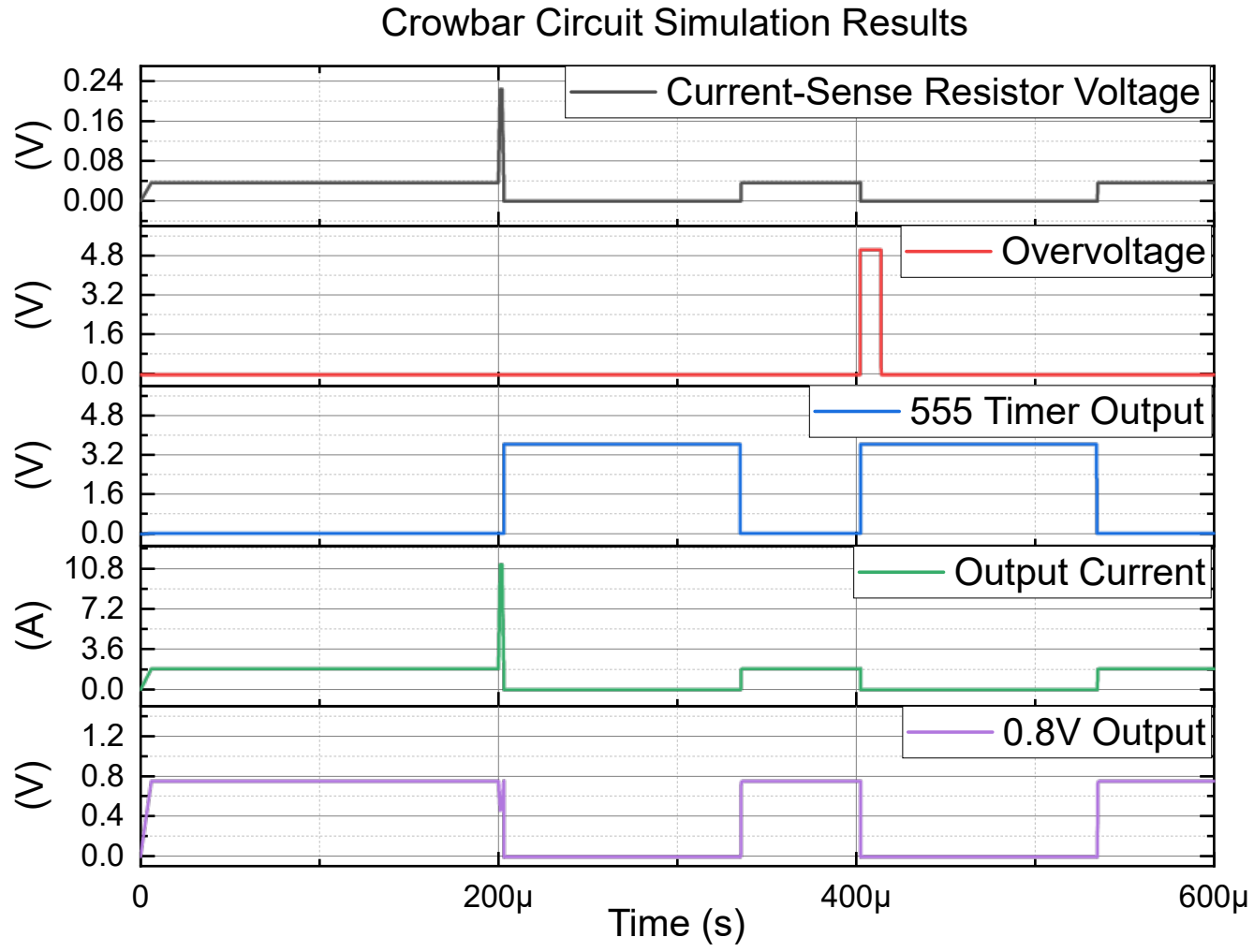

Figure 10. 0.8V single-channel crowbar circuit simulation results showing the 555 Timer output and voltage/current crowbar action when an overcurrent or overvoltaage event is detected.

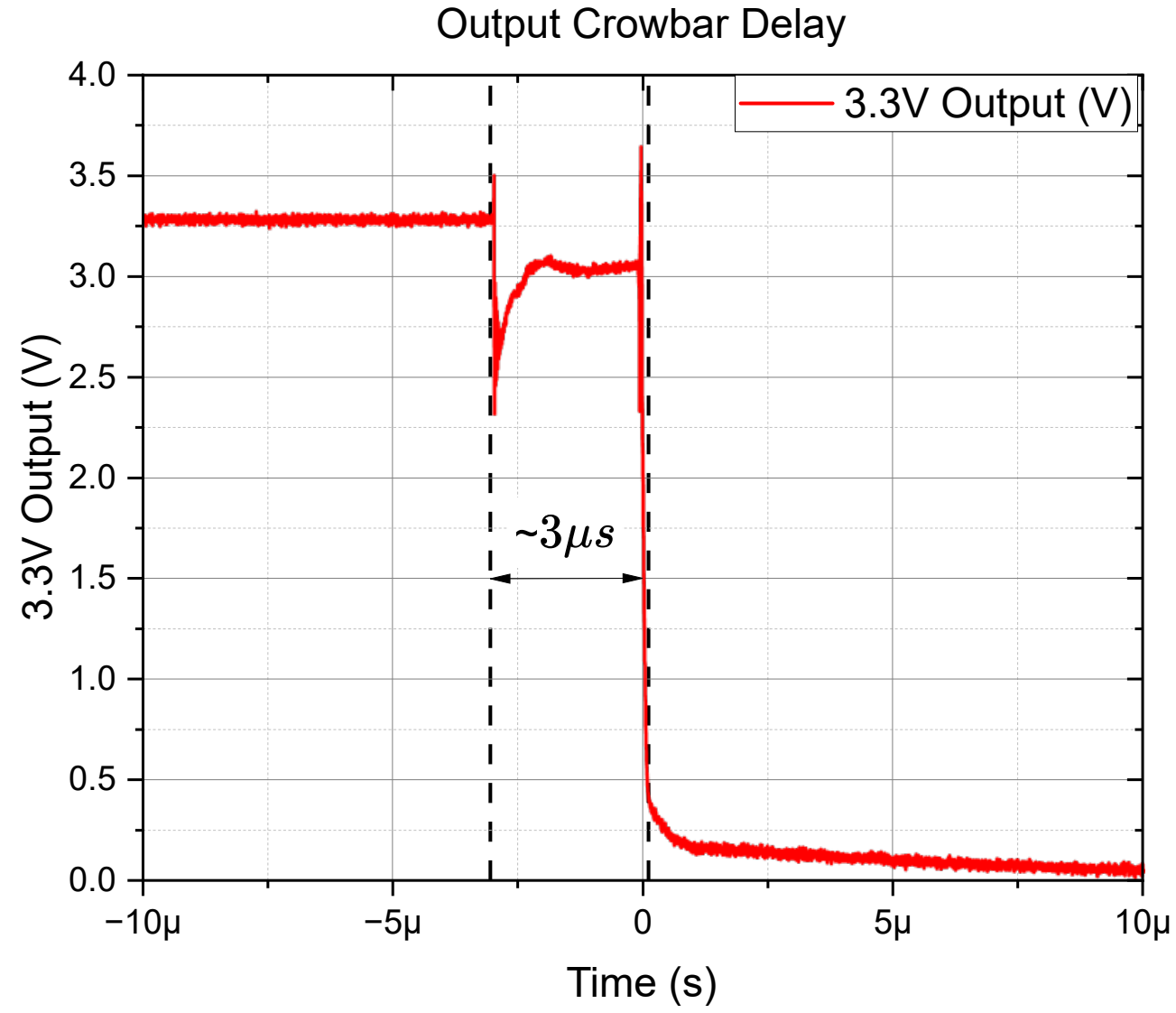

Figure 11. Test data for a 3.3V output showing the delay from the overcurrent disturbance to the crowbar circuit cutting output power.